\documentclass[11pt]{article}
\usepackage{moriond,epsfig}

\newcommand{\simlt}  {\raisebox{-.6ex}{$\stackrel{\textstyle <}{\sim}$}}

\newcommand{\jcp} {\ensuremath{J}}

\newcommand{\Uai} {U_{\alpha i}}
\newcommand{\Uajst} {U^*_{\alpha j}}
\newcommand{\Ubist} {U^*_{\beta i}}
\newcommand{\Ubj} {U_{\beta j}}
\newcommand{\LD} {L_{\Delta}}
\newcommand{\ND} {N_{\Delta}}
\newcommand{\TT} {\ensuremath{\widetilde{T}}}
\newcommand{\PT} {\ensuremath{\widetilde{P}}}

\newcommand{\mat}[9] {\left( \matrix{#1 & #2 & #3 \cr
                                     #4 & #5 & #6 \cr 
                                     #7 & #8 & #9 \cr} \right)}

\newcommand{\matTwoThr}[6] {\left( \matrix{#1 & #2 & #3 \cr
                                                                       #4 & #5 & #6 \cr} \right)}

\newcommand{\mt}{$\mu$-$\tau$}

\newcommand{\JJ} {J}
\newcommand{\FF}{{\ensuremath{\cal F}}}
\newcommand{\GG}{{\ensuremath{\cal G}}}
\newcommand{\CC}{{\ensuremath{\cal C}}}
\newcommand{\AC}{{\ensuremath{\cal A}}}
\newcommand{\BB}{{\ensuremath{\cal B}}}
\newcommand{\DD}{{\ensuremath{\cal D}}}
\newcommand{\LL}{{\ensuremath{\cal L}}}
\newcommand{\NN}{{\ensuremath{\cal N}}}

\newcommand{\Sthl}{\ensuremath{S3_{\downarrow}}}
\newcommand{\Sthn}{\ensuremath{S3_{\,\uparrow}}}

\newcommand{\bo}{\ensuremath{\bf 1}}
\newcommand{\bob}{\ensuremath{\bf \overline{1}}}
\newcommand{\Tr}{\ensuremath{{\rm Tr}}}
\newcommand{\Det}{\ensuremath{{\rm Det}}}

\newcommand{\beq}{\begin{equation}}
\newcommand{\eeq}{\end{equation}}
\newcommand{\bea}{\begin{eqnarray}}
\newcommand{\eea}{\end{eqnarray}}

\newcommand{\PP}{{\ensuremath{P}}}
\newcommand{\Lmt} {\widetilde{L^m}}
\newcommand{\Nmt} {\widetilde{N^m}}

\def\be{\begin{equation}}
\def\ee{\end{equation}}
\def\bea{\begin{eqnarray}}
\def\eea{\end{eqnarray}}

\begin{document}
\vspace*{4cm}
\title{Flavour Permutation Symmetry and Fermion Mixing \footnote{Talk given at the 43rd 
Rencontres de Moriond, La Thuile, Italy, March 2008.}
}

\author{P.F.~Harrison \footnote{Speaker.} and D.R.J. Roythorne}

\addressTwo{Department of Physics, University of Warwick, Coventry,\\
CV4 7AL, England.}

\author{W.G.~Scott}

\address{Rutherford Appleton Laboratory, Chilton, Didcot,\\
OXON, OX11 0QX, England.}

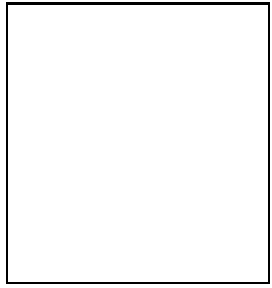
\maketitle\abstracts{
We discuss our recently proposed $\Sthl\times\Sthn$ 
flavour-permutation-symmetric mixing observables, giving expressions for 
them in terms of (moduli-squared) of the mixing matrix elements. We outline their 
successful use in providing flavour-symmetric descriptions of (non-flavour-symmetric) 
lepton mixing schemes. We develop our partially unified flavour-symmetric 
description of both quark and lepton mixings, providing testable predictions
for $CP$-violating phases in both $B$ decays and neutrino oscillations.
}

\section{Introduction}
Flavour observables, namely quark and lepton masses and mixings are neither predicted 
nor predictable in the Standard Model. Neither are they correlated with each
other in any way. However, their experimentally determined values display striking 
structure: viewed on a logarithmic scale, the fermion masses of any given non-zero charge
are approximately equi-spaced; the spectrum of quark mixing angles is described by
the Wolfenstein form,~\cite{wolfenstein} suggestive of correlations between mixing 
angles and quark masses, and the lepton mixing matrix is well-approximated by the
tri-bimaximal form.~\cite{TBM:1} These striking patterns are the modern-day equivalents 
of the regularities observed around a century ago in hydrogen emission spectra, which 
were mathematically well-described by the Rydberg formula, but nevertheless had no 
theoretical basis before the advent of quantum mechanics. While consistent with the 
Standard Model, they lie completely outside its predictive scope, and are surely 
evidence for some new physics beyond it.

In this talk, we report on our recent attempts~\cite{HRS07} to find a new description of 
fermion mixing which builds on the Standard Model and allows constraints on the mixing 
observables which make no reference to individual flavours, 
while describing mixing structures which are manifestly not flavour-symmetric,
as observed experimentally. This approach does not in itself constitute 
a complete theory of flavour mixing beyond the Standard Model, but we hope that 
it might help stimulate new developments in that direction.

\section{The Jarlskogian and Plaquette Invariance}
Jarlskog's celebrated $CP$-violating invariant,~\cite{JCP:1}  $\jcp$,
is important in the phenomenology of both quarks and leptons. As well as 
parameterising the violation of a specific symmetry, it has two other properties 
which set it apart from most other mixing observables. 
First, its value (up to its sign) is independent of any flavour labels.\footnote{We 
focus first on the leptons, although many of our considerations may be 
applied equally well to the quarks. In the leptonic case, neutrino mass eigenstate 
labels $i=1...3$ take the analogous role to the charge $-\frac{1}{3}$ quark flavour 
labels in the quark case. In this sense, we will often use the term ``flavour'' 
to include neutrino mass eigenstate labels, as well as charged lepton flavour labels.} 
Mixing observables are in general dependent on flavour labels, eg.~the 
moduli-squared of mixing matrix elements, $|\Uai|^2$, certainly depend on 
$\alpha$ and $i$. Indeed, $\jcp$ itself is often 
calculated in terms of a subset of four mixing matrix elements, 
namely those forming a given plaquette~\cite{BJ} (whose elements are defined by 
deleting the $\gamma$-row and the $k$-column \footnote{We use a cyclic 
labelling convention such that $\beta=\alpha+1$, $\gamma=\beta+1$, $j=i+1$, $k=j+1$, 
all indices evaluated mod 3.} to leave a rectangle of four elements):
\begin{eqnarray}
\jcp={\rm Im}(\Pi_{\gamma k})={\rm Im}(U_{\alpha i}U^*_{\alpha j}U^*_{\beta i}U_{\beta j}).
\label{jarlskogian}
\end{eqnarray}
However, it is well-known~\cite{JCP:1} that the value of $\jcp$ 
does not depend on the choice of plaquette (ie.~on its flavour labels, $\gamma$ and 
$k$ above) - it is 
``plaquette-invariant''. This special feature originates in the fact that \jcp\ is 
{\it flavour-symmetric}, carrying information sampled evenly across the whole 
mixing matrix. We recently pointed-out~\cite{HRS07} that in fact, {\it any} observable function 
of the mixing matrix elements, flavour-symmetrised (eg.~by summing over both 
rows and columns), and written in terms of the elements of a single plaquette 
(eg.~using unitarity constraints), will be similarly plaquette-invariant. Both its expression
in terms of mixing matrix elements, as well as its value, will be independent of 
the particular choice of plaquette.

The second exceptional property of \jcp\ is that it may be particularly 
simply related to the fermion mass (or Yukawa) matrices:
\begin{equation}
\jcp=-i\,\frac{{\rm Det}[L,N]}{2\LD \ND}
\label{jarlskogComm}
\end{equation}
where for leptons, $L$ and $N$ are the charged-lepton and neutrino mass 
matrices respectively \footnote{Throughout this paper, $L$ and $N$ are taken to be 
Hermitian, either by appropriate choice of the flavour basis for the right-handed fields, 
or as the Hermitian squares, $MM^{\dag}$, of the relevant mass or Yukawa coupling 
matrices. The symbols $m_{\alpha}$, $m_i$ generically refer to their eigenvalues in 
either case.} (in an arbitrary weak basis) and 
$\LD=(m_e-m_{\mu})(m_{\mu}-m_{\tau})(m_{\tau}-m_e)$ (with an analogous 
definition for $\ND$ in terms of neutrino masses and likewise for the 
quarks). This is useful, as, despite $\jcp$ being defined purely in terms of 
mixing observables via Eq.~(\ref{jarlskogian}), by contrast, 
Eq.~(\ref{jarlskogComm}) relates it to the mass matrices, 
which appear in the Standard Model Lagrangian.

We will discuss our recently proposed \cite{HRS07} plaquette-invariant 
(ie.~flavour-symmetric mixing) observables, which, in common with $\jcp$, 
are independent of flavour labels and can be simply related to the 
mass matrices. Again like $\jcp$, we find that our observables 
parameterise the violation of certain phenomenological symmetries which 
have already been considered 
significant~\cite{SYMMSGENS}~\cite{MUTAUSYMM}~\cite{DEMOCRACY}~\cite{BHS05} in leptonic 
mixing. In the next section, we define more precisely what we mean by flavour symmetry.

\section{The {\boldmath $\Sthl\times S3_{\uparrow}\!\!$}~Flavour Permutation Group}
The $\Sthl$ group is the group of the six possible permutations of the charged lepton
flavours and/or of the charge $-\frac{1}{3}$ quark flavours, while the $\Sthn$ group
is the group of the six possible permutations of the neutrino flavours (ie.~mass 
eigenstates) or of the charge $\frac{2}{3}$ quark flavours (the arrow subscript 
corresponds to the direction of the z-component of weak isospin of the corresponding 
left-handed fields). We consider all possible such permutations,
which together constitute the direct product $\Sthl\times\Sthn$ flavour permutation group
(FPG)~\cite{HRS07} with 36 elements.

We next consider the $P$ matrix (for ``probability'')~\cite{HSW06} of moduli-squared 
of the mixing matrix elements, eg.~for leptons:
\begin{eqnarray}
P
=\mat{|U_{e1}|^2}{|U_{e2}|^2}{|U_{e3}|^2}
     {{|U_{\mu 1}|^2}}{{|U_{\mu 2}|^2}}{{|U_{\mu 3}|^2}}
     {{|U_{\tau 1}|^2}}{{|U_{\tau 2}|^2}}{{|U_{\tau 3}|^2}}.
\label{Pmatrix}
\end{eqnarray}
It should be familiar: for quarks, semileptonic weak decay rates of hadrons are 
proportional to its elements, while for leptons, the magnitudes of neutrino oscillation 
probabilities may be written in terms of its elements.~\cite{HSW06} Moreover, the $P$ 
matrix may easily be related to the fermion mass matrices, as we will see in Section 5 
below. The $P$ matrix manifestly transforms as the natural representation of 
$\Sthl\times\Sthn$, the transformations being effected by pre- and/or post-multiplying 
by $3\times 3$ real permutation matrices.\footnote{Less obviously, any given plaquette 
of $P$ transforms as a 2-dimensional (real) irreducible representation of 
$\Sthl\times\Sthn$.}

Jarlskog's invariant $J$ is a pseudoscalar under the FPG: under even permutations, it
is invariant, while under odd permutations (eg.~single swaps of rows or columns of the
mixing matrix, or odd numbers of them), it simply changes sign. This is our prototype
Flavour Symmetric Mixing Observable (FSMO). As we commented in the previous section,
it is easy to find other similar such quantities, which, surprisingly had not 
appeared in the literature until recently.~\cite{HRS07} There are two types of singlets 
under the S3 group:
even (\bo) which remain invariant under all permutations, and odd (\bob) which flip sign 
under odd permutations. So, under the FPG, there are four types of singlet: 
{$\bf 1$}$\times${$\bf 1$}, {$\bf\overline{1}$}$\times${$\bf\overline{1}$} (like $\jcp$), 
{$\bf 1$}$\times${$\bf\overline{1}$} and {$\bf\overline{1}$}$\times${$\bf 1$}. By Flavour
Symmetric Observables (FSOs), we mean observables with any of these transformation
properties under the FPG. They may be functions of mixing matrix elements alone (FSMOs),
or functions of mass eigenvalues alone, or functions of both.

Starting with elements of $P$ and combining and (anti-)symmetrising them over flavour 
labels in various ways, 
we find that, apart from their (trivial) overall normalisation, and possibly scalar 
offsets, there are a finite number of independent FSMOs at any given order in $P$.
Enumerating them, we found that there are no non-trivial ones linear 
in $P$, while at 2nd order in $P$, there is only one each of 
{$\bf 1$}$\times${$\bf 1$}, {$\bf\overline{1}$}$\times${$\bf\overline{1}$}.
At third order, there is exactly one each of the four types of singlet, while at
higher orders in $P$, there are multiple instances of each. Recognising that we need
only four independent variables to specify the mixing, it is clearly enough to stop
at third order, up to which, the singlets are essentially uniquely defined by their 
order in $P$ and their transformation property under the FPG.

\section{Flavour-Symmetric Mixing Observables}
We introduce four FSMOs,~\cite{HRS07} uniquely defined as outlined above:
\begin{equation}
\renewcommand{\arraystretch}{1.50} 
\begin{array}{lcc}
                      &  \underline{\bo\times\bo}          & \underline{\bob\times\bob}                 \\
\underline{{\rm 2nd~Order~in}~P:} &        \GG=\frac{1}{2}\,[\,\sum_{\alpha i}(P_{\alpha i})^2-1\,] & \FF=\Det\PP \\
\underline{{\rm 3rd~Order~in}~P:} & \CC=\frac{3}{2}\,\sum_{\alpha i}[\,(P_{\alpha i})^3-(P_{\alpha i})^2\,]+1 
                      & \quad\AC=\frac{1}{18}\sum_{\gamma k}(L_{\gamma k})^3
\end{array}\label{invariants}
\end{equation}
where $L_{\gamma k}=(P_{\alpha i}+P_{\beta j}-P_{\beta i}-P_{\alpha j})$. Alternative,
but equivalent definitions in terms of the elements of a single plaquette are given 
elsewhere.~\cite{HRS07} Note that $\FF$ is only quadratic in $P$, because of the constraints 
of unitarity. We comment briefly on the normalisations and offsets we have given them. 
$\FF$ and $\AC$, being anti-symmetric,
need no offset, as they are already centred on zero, which they reach for threefold 
maximal mixing~\cite{TRIMAX} (uniquely defined by all 9 elements of the mixing 
matrix having magnitude $\frac{1}{\sqrt{3}}$). 
$\GG$ and $\CC$ are defined with offsets such that they likewise vanish 
for threefold maximal mixing. All four variables are normalised so that their maximum 
value is unity, which they attain for no mixing. In Ref.~\cite{HRS07}, 
we also give the $\bob\times\bo$ and the $\bo\times\bob\,$ FSMOs at 3rd order 
(called \BB\ and \DD\ respectively), but they will not concern us here.

The four FSMOs introduced in Eq.~\ref{invariants} are the simplest ones \footnote{They also 
treat the two weak-isospin sectors symmetrically, though this is not an essential feature.}
in terms of $P$ and are sufficient to completely specify the mixing, up to a number of 
discrete ambiguities associated with the built-in flavour symmetry. $\jcp$ is of course
not independent, and is given by $18\JJ^2=1/6 - \GG + (4/3)\,\CC - (1/2)\,\FF^2$.
In Table \ref{table:values}, we summarise their properties and values (estimated at 
90\% CL from compilations of current experimental results) for both quarks~\cite{CKMUTFIT} 
and leptons.~\cite{FITS:1}
\begin{table*}[h]
\renewcommand{\arraystretch}{1.25} 
\caption{Properties and values of flavour-symmetric mixing observables for quarks and 
leptons. The experimentally allowed ranges are estimated (90\% CL) from compilations 
of current experimental results, neglecting any correlations between the input 
quantities.\label{table:values}}
\vspace{0.4cm}
\begin{center}
\begin{tabular}{|c|c|c|c|c|c|}
\hline
  Observable  &  Order     & Symmetry:            &Theoretical& Experimental Range & Experimental Range  \\
     Name       &   in \PP\ &  $\Sthl\times\Sthn$  &   Range    & for Leptons   &  for Quarks      \\
\hline
$\FF$ 	& 2 & {$\bf\overline{1}$}$\times${$\bf\overline{1}$} &   $(-1, 1)$   	&  $(-0.14, 0.12)$  	& $(0.893, 0.896)$ \\
$\GG$ 	& 2 & {$\bf 1$}$\times${$\bf 1$}                &   $(0, 1)$   		&  $(0.15, 0.23)$  	& $(0.898, 0.901)$ \\
$\AC$ 	& 3 & {$\bf\overline{1}$}$\times${$\bf\overline{1}$}&   $(-1, 1)$   	&  $(- 0.065, 0.052)$ & $(0.848, 0.852)$ \\
$\CC$ 	& 3 & {$\bf 1$}$\times${$\bf 1$}&   $(-\frac{1}{27}, 1)$  		&  $(-0.005, 0.057)$  & $(0.848, 0.852)$ \\
\hline
\end{tabular}
\end{center}
\end{table*}

\section{Flavour-Symmetric Mixing Observables in Terms of Mass Matrices}
Equation (\ref{jarlskogComm}) gives $\jcp$, our prototype FSMO, in terms of the 
fermion mass matrices, which in turn are proportional to the matrices of Yukawa 
couplings which appear in the Standard Model Lagrangian. In this section, we show 
how to write the FSMOs of Section 4 above also in terms of the mass matrices.
It is useful to define a reduced $P$ matrix:
\beq
\PT=P-D
\label{PT}
\eeq
where $D$ is the $3\times 3$ democratic matrix with all 9 elements equal to $\frac{1}{3}$.
We also define the reduced (ie. traceless) powers of the fermion mass matrices: 
$\Lmt:=L^m-\frac{1}{3}\Tr(L^m)$ (similarly for $\Nmt$), in terms of which, we can define
the $2\times 2$ matrix of weak basis-invariants:
\beq
\TT_{mn}:={\rm Tr}(\widetilde{L^m}\widetilde{N^n}),\quad m,n=1,2.
\eeq
For known lepton masses, \TT\ is completely equivalent to \PP. In fact, it is 
straightforward to show that $\PT$ is a mass-moment transform of $\TT$:
\beq
\PT=\widetilde{M_{\ell}}^T\cdot \TT\cdot \widetilde{M_{\nu}}
\label{PTexpan}
\eeq
where
\bea
\widetilde{M_{\ell}}=\frac{1}{\LD}
\matTwoThr{m_{\mu}^2-m_{\tau}^2}{m_{\tau}^2-m_e^2}{m_{e}^2-m_{\mu}^2}
    	  {m_{\mu}-m_{\tau}}{m_{\tau}-m_e}{m_{e}-m_{\mu}},
\label{PtildeFromTtilde}
\eea
with an analogous definition for $\widetilde{M_{\nu}}$ (the inverse transform is 
easily obtained).

Starting from Eq.~(\ref{invariants}) and substituting for $P$ from Eqs.~(\ref{PT})
and (\ref{PTexpan}), we find that:
\beq
\FF\equiv{\rm Det}\,P=3\frac{{\rm Det}\,\TT}{\LD\ND};
\qquad\quad \left[{\rm cf.~Eq.~(\ref{jarlskogComm}):}~\jcp=-i\,\frac{{\rm Det}[L,N]}{2\LD \ND}\right]
\eeq
\beq
\GG=\frac{\TT_{mn}\,\TT_{pq}\,\LL^{mp}\,\NN^{nq}}{(\LD\ND)^2};
\qquad\CC,\AC=\frac{\TT_{mn}\,\TT_{pq}\,\TT_{rs}\,\LL_{\CC,\AC}^{(mpr)}\,\NN_{\CC,\AC}^{(nqs)}}{(\LD\ND)^{n_{\CC,\AC}}},
\eeq
where the $\LL$ ($\NN$) are simple functions of traces of $\Lmt$ ($\Nmt$), given 
in Ref.~\cite{HRS07}, and $n_{\CC}~(n_{\AC})=2(3)$.

\section{Application 1: Flavour-Symmetric Descriptions of Leptonic Mixing}
The tribimaximal mixing~\cite{TBM:1} ansatz for the MNS lepton mixing matrix:
\bea
U\simeq \left( \matrix{
		      -2/\sqrt{6}  &  1/\sqrt{3} & \,\,\,0   \cr
 	             \,\,\,\, 1/\sqrt{6}  &  1/\sqrt{3} & \,\,\,\,\, 1/\sqrt{2}  \cr
		     \,\,\,\, 1/\sqrt{6}  &  1/\sqrt{3} &  -1/\sqrt{2} \cr
} \right)
\eea
is compatible with all confirmed leptonic mixing measurements
from neutrino oscillation experiments, and may be considered a useful
leading-order approximation to the data. It is defined by three phenomenological 
symmetries:~\cite{SYMMSGENS} $CP$ symmetry, \mt-reflection symmetry and Democracy, 
which may each be expressed (flavour-symmetrically) in terms of our FSMOs. For example, 
as is well known, the zero in the $U_{e3}$ position, if exact, ensures that no 
$CP$ violation can arise from the mixing matrix. $CP$ symmetry is thus represented 
simply by $\jcp=0$ (which is a necessary, but not sufficient condition for a single 
zero in the mixing matrix, see Section 7 below). \mt-reflection symmetry~\cite{MUTAUSYMM}
means that corresponding elements in the $\mu$ and $\tau$ rows have equal moduli:
$|U_{\mu i}|=|U_{\tau i}|,~\forall i$, and this implies the two flavour-symmetric
constraints:
\beq
\FF=\AC=0
\label{mtSymm}
\eeq
(flavour symmetry means that although these two constraints imply just such a set of 
equalities, they do not define {\it which} pair of rows or columns are constrained). 
Democracy~\cite{DEMOCRACY}~\cite{BHS05} ensures that one row or column is trimaximally 
mixed, ie.~has the form $\frac{1}{\sqrt{3}}(1,1,1)^{(T)}$, as is the case for the 
$\nu_2$ column in tribimaximal mixing. Democracy is ensured flavour-symmetrically 
by the two constraints:
\beq
\FF=\CC=0.
\label{democracy}
\eeq
Taking all three symmetries, tribimaximal mixing (or one of its trivial
permutations) is ensured by the complete set of constraints $\FF=\CC=\AC=\jcp=0$, which
may be written as the single flavour-symmetric condition:
\beq
\FF^2+\CC^2+\AC^2+\jcp^2=0.
\label{tbmConstraint}
\eeq

Tribimaximal mixing is manifestly not flavour symmetric. The flavour-symmetry of our
constraint, Eq.~(\ref{tbmConstraint}), is spontaneously broken by its tribimaximal 
solutions. The symmetry is manifested by the existence of a complete set of solutions 
of the generalised tribimaximal form, each related to the other by a member of the 
flavour permutation group.

Of course, generalisations of the tribimaximal form~\cite{SYMMSGENS} possessing 
subsets of its three symmetries may be similarly defined, and their corresponding 
flavour-symmetric constraints may be obtained by analogy to the above. These, and
those of other special mixing forms~\cite{AF}~\cite{BIMAX} are tabulated in 
Ref.~\cite{HRS07}.

\section{Application 2: A Partially Unified, Flavour-Symmetric Description of Quark 
and Lepton Mixings}
A unified understanding of quark and lepton mixings is highly desirable. This is
difficult because their mixing matrices have starkly different forms: 
the quark mixing matrix is characterised by small mixing angles,~\cite{CKMUTFIT} 
while the lepton mixing matrix is characterised mostly by large ones.~\cite{FITS:1} 
Many authors have ascribed this difference to the effect of the heavy majorana mass 
matrix in the leptonic case, via the see-saw mechanism.~\cite{seesaw} Notwithstanding 
the attractiveness of this explanation, it is clearly still worthwhile to ask if there 
are any features of the respective mixings which the quark and lepton sectors have in
common.

Neutrino oscillation data~\cite{FITS:1} require that $|U_{e3}|^2\,\simlt\,0.05$, 
significantly less than the other MNS matrix elements-squared. At least one {\em small} 
mixing element is hence a common feature of both quark and lepton mixing matrices. 
We are thus led first to ask the question:~``what is the flavour-symmetric condition 
for at least one zero element in the mixing matrix?'' We should perhaps anticipate 
two constraints, as the condition implies that both real and imaginary parts 
vanish. A zero mixing element implies $CP$ conservation, so that $\jcp=0$. 
A clue to the second constraint is that with \mt-reflection symmetry, 
$\jcp=0$ ensures a zero somewhere in the $\nu_e$ row of the MNS matrix. However, 
\mt-reflection symmetry implies two more constraints, Eq.~(\ref{mtSymm}).

In order to find a single additional constraint we consider the 
$K$ matrix~\cite{Kmatrix}~\cite{HSW06} with elements:
\beq
K_{\gamma k}={\rm Re}(\Uai\Uajst\Ubist\Ubj),
\eeq
which is the $CP$-conserving analogue of $\jcp$ (cf.~the definition of \jcp, 
Eq.~(\ref{jarlskogian})).
$K$ should be familiar: in the leptonic case, its elements are often
used to write the magnitudes of the oscilliatory terms in neutrino appearance 
probabilities;~\cite{HSW06} in the quark case, its elements are just the CKM factors 
of the $CP$-conserving parts of the interference terms in penguin-dominated decay rates. 
A single zero in the mixing matrix leads to four zeroes in a plaquette of $K$ and
this clearly implies:
\beq
\Det\,K=0,
\label{constraint2}
\eeq
which is our sufficient second condition, along with $\jcp=0$.~\footnote{The two
conditions may even be expressed as one, noting that the product of all nine
elements of $P$ is given by 
$\frac{1}{144}\prod_{\alpha i} P_{\alpha i}=(\Det\,K)^2+\jcp^2(2\jcp^2+{\cal{R}})^2$, 
which is zero iff $\Det\, K=0$ and $\jcp=0$ (as ${\cal{R}}>0$, as long as $\jcp\neq 0$).} 
We note that Eq.~(\ref{constraint2}) can easily be cast in terms of our complete 
set of FSMOs, since $54\,\Det\,K\equiv 2\AC + \FF(\FF^2 - 2\CC - 1)$. Hence, 
\mt-reflection symmetry, Eq.~(\ref{mtSymm}), is a special case of Eq.~(\ref{constraint2}).

Experimentally, there is no exactly zero element in the CKM matrix, so that $\Det\,K=0$ 
{\em and} $\jcp=0$ cannot {\em both} be exact for quarks. Moreover, for leptons,
despite there being no experimental lower limit for $|U_{e3}|$, there is no reason to 
suppose that the MNS matrix has an exact zero either. In order to ensure a 
small, but non-zero element in the mixing matrices, we need to consider a 
modest relaxation of either condition, or of both.
For quarks, we know from experiment that $CP$ is slightly violated, with~\cite{CKMUTFIT} 
$|\jcp_q/\jcp_{max}|\simeq 3\times 10^{-4}$, while~\footnote{We note that 
$\jcp_{max}=\frac{1}{6\sqrt{3}}\simeq 0.1$ and $(\Det\,K)_{max}=\frac{2^6}{3^9}\simeq 0.0033$.} 
for leptons, fits to oscillation 
data~\cite{FITS:1} imply a fairly loose upper bound on their $CP$ violation: 
$|\jcp_{\ell}/\jcp_{max}|\,\simlt\,0.33$. Turning to $\Det\,K$, we find that for quarks,
$|\Det\,K_q/(\Det\,K)_{max}|\,\simlt\, 3\times 10^{-7}$, while for leptons, 
$|\Det\,K_{\ell}/(\Det\,K)_{max}|\,\simlt\, 0.6$ (the precision of lepton mixing data 
does not yet allow a strong constraint). However, there is no experimental lower limit for 
$|\Det\,K|$ for quarks or for leptons, each being compatible with zero, so that it is 
sufficient to relax only the condition on $\jcp$.

We are thus led to conjecture that for both quarks and leptons:
\beq
\Det\,K=0;\quad |\jcp/\jcp_{max}|={\rm small}
\label{conjecture}
\eeq
(it is not implied that the small quantity necessarily has the same value in both 
sectors). Equation (\ref{conjecture}) is a unified and flavour-symmetric, partial 
description of both lepton and quark mixing matrices, being associated with
the existence of at least one small element in each mixing matrix, $U_{e3}$ and $V_{ub}$ 
respectively (it is partial in the sense that only two degrees of freedom are constrained
for each matrix). However, in the case that $\jcp$ is not exactly zero, the condition 
$\Det\,K=0$ also implies that in the limit, as $\jcp\rightarrow 0$, there is at 
least one unitarity triangle angle which 
$\rightarrow 90^{\circ}$. This is rather obvious in the \mt-symmetry case, but is 
less obvious more generally. While the flavour symmetry prevents an a priori prediction
of {\it which} angle is $\simeq 90^{\circ}$, we know from experiment~\cite{CKMUTFIT} that 
for quarks, $\alpha\simeq 90^{\circ}$. A detailed calculation shows that our conjecture, 
Eq.~(\ref{conjecture}), predicts, in terms of Wolfenstein parameters:~\cite{wolfenstein}
\beq
(90^{\circ}-\alpha)=\overline\eta\lambda^2=1^{\circ}\pm0.2^{\circ}
\label{qPrediction}
\eeq
at leading order in small quantities, to be compared with its current experimental 
determination:~\cite{CKMUTFIT}
\beq
(90^{\circ}-\alpha)=0^{\circ+3^{\circ}}_{\,\,\,-7^{\circ}}.
\eeq
It will be interesting to test Eq.~(\ref{qPrediction}) more precisely in future 
experiments with $B$ mesons, in particular, at LHCb and at a possible future Super 
Flavour Factory. For leptons, experiment tells us not only that it is the $U_{e3}$ MNS 
matrix element which is small but also that only the unitarity triangle 
angles~\footnote{We use the
nomenclature of unitarity triangle angles we defined in reference [46] of Ref.~\cite{BHS05}.} 
$\phi_{\mu 1}$ or $\phi_{\tau 1}$ can be close to $90^{\circ}$. 
Then Eq.~(\ref{conjecture}) implies that:
\beq
|90^{\circ}-\delta|=2\sqrt{2}\,\sin{\theta_{13}}\,\sin{(\theta_{23}-\frac{\pi}{4})}\,\simlt\, 4^{\circ}
\eeq
at leading order in small quantities (we use the PDG convention here). It thus requires 
a large $CP$-violating phase in the MNS matrix, which is promising for the discovery of 
leptonic $CP$ violation at eg.~a future Neutrino Factory.

\section{Discussion and Conclusions}
Given that our flavour-symmetric variables are defined (essentially) uniquely by their
flavour symmetry properties and by their order in $P$, it is remarkable that the 
leptonic data may be described simply by the constraints $\FF=\AC=\CC=\jcp=0$.
This is suggestive that these variables may be fundamental in some way.
It is furthermore tantalising that the smallness of one element in each mixing matrix, 
the approximate \mt-symmetry in lepton mixing and the existence of a right
unitarity triangle may all be related to each other, through our simple partially-unified 
constraint, Eq.~(\ref{conjecture}). The precision of the resulting 
prediction, Eq.~(\ref{qPrediction}), motivates more sensitive tests at future
$B$ physics facilities, while the synergy with tests at a neutrino factory is manifest.

All elements of the Standard Model, apart from the Yukawa couplings of the fermions 
to the Higgs, treat each fermion of any given charge on an equal footing - they are 
already flavour-symmetric. The Yukawa couplings, on the other 
hand, depend on flavour in such a way that each flavour has unique mass and mixing matrix 
elements. Using our flavour-symmetric observables, or combinations of them appropriately
chosen, we have shown how it is also possible to specify the flavour-dependent mixings 
in a flavour-independent way.~\footnote{We illustrated another variant of this in 
Ref.~\cite{EXTREMISATION}.} This recovers flavour symmetry at the level of
the mixing description, the symmetry being broken only spontaneously by its solutions, 
which define and differentiate the flavours in terms of their mixings.

\section*{Acknowledgments}
PFH thanks the organisers of the 43rd Rencontres de Moriond for organising a very 
stimulating conference. PFH also acknowledges the hospitality of the Centre for 
Fundamental Physics (CfFP) at the STFC Rutherford Appleton Laboratory. This work 
was supported by the UK Science and Technology Facilities Council (STFC).

\end{document}